\documentclass[10pt,letterpaper]{article}  
\usepackage{opex3}

\DeclareGraphicsExtensions{.eps}
\graphicspath{{./figures/}}

\usepackage{float}
\usepackage{xspace,amsmath,amssymb}
\usepackage{array}
\usepackage{cite}

\newcommand*{\indfoc}{{\rm foc}}
\newcommand*{\indcal}{{\rm cal}}
\newcommand*{\inddm}{{\rm DM}}
\newcommand*{\inddiv}{{\rm div}}
\newcommand*{\inddh}{{\rm DH}}
\newcommand*{\inddet}{{\rm det}}

\newcommand*{\tf}{\mathcal{F}}
\newcommand*{\volt}{\text{v}}
\newcommand*{\En}{\mathcal{E}}
\newcommand*{\regul}{\mathcal{R}}
\newcommand*{\lovD}{\frac{\lambda}{D}}
\newcommand*{\lovDt}{\lambda/D}
\newcommand*{\reg}{\mathcal{R}}
\newcommand*{\grad}[2]{\frac{\partial{#1}}{\partial{#2}}}
\newcommand*{\norm}[1]{\left\|{#1}\right\|}
\newcommand*{\module}[1]{\left|{#1}\right|}

\newcommand*{\itemdiam}[1]{
\begin{itemize}
\renewcommand{\labelitemi}{$\diamond$}
\setlength\itemsep{-0.18in}
{#1}
\end{itemize}
}

\newcommand*{\review}[1]{#1}

\begin{document}

\title{High-order myopic coronagraphic phase diversity (COFFEE) for wave-front
  control in high-contrast imaging systems.}

\author{B. Paul,$^{1,2,3,*}$ L. M. Mugnier,$^{1,3}$ J.-F. Sauvage$^{1,3}$ and
  K. Dohlen $^{2,3}$}
\address{$^1$Onera - The French Aerospace Lab, \\ F-92322 Chatillon, France\\
$^2$Aix Marseille Universit\'e, CNRS, LAM (Laboratoire
  d'Astrophysique de Marseille), 
\\ UMR 7326, 13388, Marseille, France\\
$^3$Groupement d'int\'er\^et scientifique PHASE (Partenariat Haute
r\'esolution Angulaire Sol et Espace) between Onera, Observatoire de Paris, 
CNRS, Universit\'e Diderot, Laboratoire d'Astrophysique de Marseille and
Institut de Plan\'etologie et d'Astrophysique de Grenoble, France}
\email{baptiste.paul@onera.fr}

\begin{abstract}
  The estimation and compensation of quasi-static aberrations is mandatory to
  reach the ultimate performance of high-contrast imaging systems. COFFEE is a
  focal plane wave-front \review{sensing method} that consists in the
  extension of phase diversity to high-contrast imaging systems. Based on a
  Bayesian approach, it estimates the quasi-static aberrations from two focal
  plane images recorded from the scientific camera itself. In this paper, we
  present COFFEE's extension which allows an estimation of low and high order
  aberrations with nanometric precision for any coronagraphic device. The
  performance is evaluated by realistic simulations, performed in the SPHERE
  instrument framework. We develop a myopic estimation that allows us to take
  into account an imperfect knowledge on the used diversity phase. Lastly, we
  evaluate COFFEE's performance in a compensation process, to optimize the
  contrast on the detector, \review{and show it allows one to reach the
    $10^{-6}$ contrast required by SPHERE at a few resolution elements from
    the star}. Notably, we present a non-linear energy minimization method
  which can be used to reach very high contrast levels \review{(better than $10^7$
    in a SPHERE-like context)}.
\end{abstract}


\section{Introduction}

Exoplanet imaging is one of the most challenging areas in today's astronomy.
The observation of an extremely faint object (the planet) very close to a
bright source (the host star) requires the use of an extreme adaptive optics
(XAO) system coupled with a high-contrast imaging technique such as
coronagraphy. The current generation of instruments dedicated to exoplanets
direct imaging (SPHERE on the VLT \cite{Beuzit-p-07}, GPI on Gemini \review{South}
\cite{gpi}, Subaru SCExAO \cite{spie_2010_oguyon} and Palomar P1640
\cite{pasp_11_hinkley}) aim at detecting massive gaseous planets $10^{-6}$ to
$10^{-7}$ times fainter than their host star. In the future, high-contrast imaging
instruments on ground based or space based telescopes will perform observation
of Earth-like planets, $10^{-9}$ to $10^{-10}$ times fainter than their host star.

The ultimate limitation of a high-contrast imaging instrument lies in its
quasi-static aberrations\review{, which originate in imperfections of the
  optical system such as misalignment or optical surface polishing error}.
These aberrations, \review{when unseen and thus} uncorrected by the AO loop,
create long-lived speckles on the detector plane \cite{apj_07_soummer},
limiting the achievable contrast. \review{Besides, unlike the signal which
  originates in the residual turbulence (averaged in a long exposure image),
  these speckles can easily be mistaken for a planet}. Thus, to reach the
ultimate performance of the imaging system, one must be able to compensate for
these aberrations. \review{To perform such a
  compensation, SPHERE relies on phase diversity \cite{Sauvage-p-12} to
  reach a contrast of $10^{-6}$, whereas GPI relies on an interferometric
  approach \cite{gpi_wallace_2010} for an aimed contrast of $10^{-7}$ on the
  detector.}

Several techniques dedicated to this compensation have been proposed. Closed
loop methods, which assume small aberrations (\cite{speckle_nulling, efc,
  spie_2007_jtrauger, spie_10_sthomas, spie_2012_pbaudoz}), estimate the
electric field in the detector plane using at least three images. The
technique proposed by Baudoz \textit{et al.} \cite{scc} relies on a modification of the
imaging system, but requires only one image. \review{We note that this
  approach, based on the analysis of fringed speckles, requires a $\sqrt{2}$
  oversampling of the coronagraphic images to properly sample the interference
  fringes.} These techniques aim at minimizing the energy in a chosen area
(``Dark Hole''), leading to a contrast optimization on the detector in a
closed loop process.

The focal plane wave-front sensor we have proposed \cite{Sauvage-a-12}, called
COFFEE (for COronagraphic Focal-plane wave-Front Estimation for Exoplanet
detection), requires only two focal-plane images to estimate the aberrations
both upstream and downstream of the coronagraph without any modification of
the coronagraphic imaging system or assuming small aberrations. In a previous
study \cite{Paul-a-13}, we presented COFFEE's early performance and
limitations, detailed below, as well as its sensitivity to a realistic
experimental environment. In this paper, we present a \review{high order
  extension of COFFEE and its performance evaluation in a compensation
  process, in the framework of the quasi-static aberration calibration of a
  ground-based instrument such as SPHERE. Section \ref{est} presents the
  modifications that allows COFFEE to overcome its previous limitations
  \cite{Paul-a-13}, allowing an estimation of high order aberrations with
  nanometric precision for any coronagraphic device. Section \ref{perf}
  presents the noise sensitivity of this extended version of COFFEE using
  realistic SPHERE-like simulations. The choice of a suitable diversity phase
  to use is also discussed in this section. Knowing that in a real system,
  such a diversity phase will not be perfectly introduced, we present, in
  Section \ref{myop}, an original approach, hereafter called ''myopic'', which
  significantly} improves COFFEE's robustness to an imprecise knowledge of the
diversity phase. Finally, in Section \ref{comp}, we describe the different
compensation processes which can be used with COFFEE. In particular, we
present a method of energy minimization in the detector plane that allows the
creation of a Dark Hole without any small aberration assumption.
\review{Unlike other energy minimization methods \cite{speckle_nulling, efc,
    spie_2007_jtrauger, spie_10_sthomas, spie_2012_pbaudoz}, the one we
  propose does not rely on the calibration of an interaction matrix, which is
  sensitive to the position of the coronagraphic image on the detector. This
  new dark hole method can thus be used on any high contrast
  instrument without a repetitive dedicated calibration step.\\
  In the SPHERE baseline design, quasi-static aberrations are measured with
  conventional phase diversity \cite{Sauvage-p-12} (no coronagraph), which is
  unable to sense high-order aberrations to a nanometric level. As a result,
  this high contrast imaging instrument performance will be limited by
  high-order phase aberrations and not by amplitude aberrations (amplitude
  variations in a pupil plane). Consequently, the latter are not considered in
  the simulations presented herein and are currently not estimated by COFFEE.}

\section{Aberrations estimation with COFFEE}
\label{est}

\review{COFFEE is based on an inverse problem approach: it estimates the
  aberrations both upstream and downstream of the coronagraph using two
  focal-plane images that differ from a known aberration.} As described in
\cite{Paul-a-13}, the two main error sources of the first version of COFFEE
have been shown both by simulations and experimentally to be aliasing and
modelling error. The former was due to the use of a Zernike basis and
prevented COFFEE from estimating high order aberrations. Moreover, the
estimation of these high order aberrations is mandatory to optimize the
contrast in the detector far from the optical axis (between a few $\lovDt$ and
$20\ \lovDt$ in the case of the SPHERE instrument).

The latter was originating in the image formation model used by
COFFEE. The estimations were indeed performed using a perfect coronagraph
model, and thus limited by a model error. In practice, COFFEE's use was limited
to the apodized Roddier \& Roddier coronagraph.

In this Section, we present COFFEE's \review{modifications that allows to get
  rid of these two limitations}. Section \ref{est_expr} describes the
modification of the maximum \textit{a posteriori} (MAP) approach on which
COFFEE is based, which includes a modification of the basis used for the
aberration estimation, now composed of pupil indicator functions (pixels).
Such a basis, used with a dedicated regularization metric described in Section
\ref{est_reg}, allows COFFEE to estimate high-order aberrations. Besides,
thanks to the modification of the imaging model, described in Section
\ref{est_img}, COFFEE is now able to perform the estimation for any
coronagraphic device.

\subsection{Criterion expression}
\label{est_expr}

Most of the notations of this article are coherent with \cite{Sauvage-a-12,
  Paul-a-13}. We consider a coronagraphic imaging system made of four
successive planes denoted by A (circular entrance pupil of diameter $D_u$), B
(coronagraphic focal plane), C (Lyot Stop), and D (detector plane). The
optical aberrations are considered as static and introduced in pupil planes A
and C. The coronagraphic device is composed of a focal plane mask located in
plane B and a Lyot Stop in plane C. No particular assumption is made on the
pupil shape or intensity, which can be calibrated using data recorded from the
instrument. \review{We note that this model does not consider out of plane
  aberrations and the corresponding amplitude aberrations (which originate in
  Fresnel effect), but as said previously, we consider in this paper the
  case of a ground based instrument (such as SPHERE) limited by phase
  aberrations.}

COFFEE requires only two images $i_c^\indfoc$ and $i_c^\inddiv$
recorded on the detector (plane D) that, as in phase diversity, differ from a
known aberration $\phi_\inddiv$, to estimate aberrations both upstream
($\phi_u$) and downstream ($\phi_d$) of the coronagraph.

In this paper, we consider the case of the instrument calibration, assumed to
be performed at high signal to noise ratio (SNR) value, with a monochromatic
source, emitted from a single-mode laser fiber. Since the impact of the
source finite size on the estimation is not significant on the aberrations
estimation \cite{Paul-a-13}, we consider here that this calibration is
performed with an unresolved object, and use the following imaging model :

\begin{equation}\label{eq_im_model}
\begin{aligned}
i_c^\indfoc&=\alpha_\indfoc h_\inddet \star h_c(\phi_u,\phi_d)+n_\indfoc+\beta_\indfoc\\
i_c^\inddiv&=\alpha_\inddiv h_\inddet \star h_c(\phi_u+\phi_\inddiv,\phi_d)+n_\inddiv+\beta_\inddiv
\end{aligned}
\end{equation}

where $\alpha_p$ is the incoming flux ($p$ is for ``foc'' or ``div''), $h_c$
the coronagraphic \review{on axis} ``point spread function'' (PSF) of the
instrument (which is the response of a coronagraphic imaging system to a point
source), $h_\inddet$ the known detector PSF, $n_\indfoc$ and $n_\inddiv$ are
the measurement noises and comprise both detector and photon noises, $\beta_p$
is a unknown uniform background (offset), and $\star$ denotes the discrete
convolution operation.

COFFEE is based on a maximum \textit{a posteriori} (MAP) approach: it
estimates the aberrations $\phi_u$ and $\phi_d$ as well as the fluxes
$\alpha=[\alpha_\indfoc,\alpha_\inddiv]$, and the backgrounds
$\beta=[\beta_\indfoc,\beta_\inddiv]$ that \review{maximize the posterior
  likelihood $p(\alpha, \beta, \phi_u, \phi_d | i_c^{\text{foc}},
  i_c^{\text{div}})$ of the data. For practical issues, it is more convenient
  (and equivalent) to minimize the opposite of the logarithm of the posterior
  likelihood, or neg-log-likelihood} $J(\alpha, \beta, \phi_u, \phi_d
)=-\text{ln}[p(\alpha, \beta, \phi_u, \phi_d | i_c^{\text{foc}},
i_c^{\text{div}})]$ which includes regularization terms $\regul(\phi_u)$ and
$\regul(\phi_d)$ designed to enforce smoothness of the sought phases:
\begin{equation}
(\hat{\alpha}, \hat{\beta}, \hat{\phi}_u, \hat{\phi}_d) = 
\underset{\alpha, \beta, \phi_u, \phi_d}{\arg \min}[J(\alpha, \beta, \phi_u, \phi_d )]
\end{equation}
where
\begin{equation}\label{eq-pb-inverse}
\begin{aligned}
J(\alpha, \beta, \phi_u, \phi_d)&=\frac{1}{2} \norm{\frac{i_c^{\text{foc}} 
- (\alpha_\indfoc h_\inddet\star h_c(\phi_u,\phi_d)+\beta_\indfoc)}{\sigma_n^{\text{foc}}}}^2 \\
&+\frac{1}{2}\norm{\frac{i_c^{\text{div}} 
- (\alpha_\inddiv h_\inddet\star h_c(\phi_u+\phi_\inddiv,\phi_d)+\beta_\inddiv)}{\sigma_n^{\text{div}}}}^2\\
&+\reg(\phi_u) + \reg(\phi_d)
\end{aligned}
\end{equation}
$\norm{x}^2$ denotes the sum of squared pixel values of map $x$,
$\sigma_n^{\text{foc}}$, and $\sigma_{n}^{\text{div}}$ are the noise standard
deviation maps of each image. The corresponding variances can be computed as a
sum of the photon and detector noise variances. The former can be estimated as
the image itself thresholded to positive values, and the latter can be
calibrated prior to the estimation.

Any aberration $\phi$ is expanded on a basis $\{b_m\}$. In \cite{Paul-a-13},
we showed that the use of a truncated Zernike basis for the reconstruction led
to a strong aliasing error, let alone the inability to estimate high
frequency aberrations. In this paper, the phase is expanded on pixel
indicator functions in the pupil plane: $\phi=\sum_m \phi^mb_m$ (with $\phi^m$ the
value of the $m-$th pixel in the pupil). Such a basis, used with the proper
regularization metrics, will allow COFFEE to estimate high order aberrations
and strongly reduce the aliasing error, as shown in the following.

The minimization of metric $J(\alpha, \beta, \phi_u, \phi_d)$ of
Eq.~\eqref{eq-pb-inverse} is performed by means of a limited memory variable
metric (BFGS) method (\cite{numerical_recipes, Thiebaut-p-02}), which is a
fast quasi-Newton type minimization method. It uses \review{the analytical
  expression of gradients $\grad{J}{\phi_u}$, $\grad{J}{\phi_d}$,
  $\grad{J}{\alpha}$ and $\grad{J}{\beta}$, which we have calculated}, to
estimate $\phi_u$, $\phi_d$,
$\alpha$ and $\beta$ (Implementation details can be found in Appendix \ref{app_grad}).\\
Sauvage \textit{et al.} \cite{Sauvage-a-12} established that a suitable diversity phase
$\phi_\inddiv$ for COFFEE was a mix of defocus and astigmatism:
$\phi_\inddiv=a_4^\inddiv Z_4 + a_5^\inddiv Z_5$ with
$a_4^\inddiv=a_5^\inddiv=0.8\text{ rad RMS}$, introduced upstream of the
coronagraph. In this paper, such a diversity phase will be used for a start;
the optimal phase diversity to be used with COFFEE will be discussed later.

\subsection{Regularization metric}
\label{est_reg}

The use of a pixel basis for the phase reconstruction is required for COFFEE
to estimate high order aberrations. However, this leads to a large number of
unknowns, which in turn calls for a regularization metric in order to reduce
the noise sensitivity. We chose a regularization metric that is based on the
available \textit{a priori} knowledge on the quasi-static aberrations. Indeed,
they can be reasonably assumed to be Gaussian, homogeneous and thus endowed
with a power spectral density (PSD) $S_{\phi_k}$ (where k stands for $u$
(upstream) or $d$ (downstream)), which is usually assumed to follow a power law:
\begin{equation}\label{eq_psd}
\left\{\begin{matrix}
S_{\phi_k} \propto \frac{1}{\nu^{n}}\\
\langle \phi_k \rangle=0
\end{matrix}\right.
\end{equation}
with $\nu$ the spatial frequency and $\langle \phi_k \rangle$ the mean of $\phi_k$. The
regularization term $\reg(\phi_k)$ can thus be written as:
\begin{equation}\label{eq_reg}
\reg(\phi_k)=\frac{1}{2}\sum_\nu\frac{\module{\tf[\phi_k](\nu)}^2}{S_{\phi_k}(\nu)},
\end{equation}
where $\tf$ represent the Fourier transform operation. In order to be able to
cope with any pupil shape, we implement this metric in direct space rather
than Fourier space as follows; for n=2 in Eq.~\eqref{eq_psd}, we obtain:
\begin{equation}\label{eq_regf2}
\reg(\phi_k)=\frac{\mu_k}{2}\norm{\nabla\phi_k(r)}^2\text{.}
\end{equation}   
And for n=4 in Eq.~\eqref{eq_psd},
\begin{equation}\label{eq_regf4}
\reg(\phi_k)=\frac{\mu_k}{2}\norm{\Delta\phi_k(r)}^2\text{.}
\end{equation}   
Here, $r$ denotes the pupil plane position which will be omitted in the
following for the sake of simplicity. $\nabla$ and $\Delta$ represent the
gradient and the Laplacian operators, respectively. The balance parameter
$\mu_k$ will be called ``hyperparameter'' hereafter. For both cases,
derivatives $\nabla\phi_k$ and $\Delta\phi_k$ are computed as finite
differences between neighboring points, and summations are limited to points
whose computation requires only pixels inside the pupil.
In this paper, we consider a PSD decrease as $1/\nu^2$ ($n=1$ in
Eq.~\eqref{eq_psd}), which corresponds to a classical assumption for
\review{optical surface polishing errors, according to K. Dohlen \textit{et
    al.}\cite{Dohlen-p-11}, who measured and characterized the PSD of the
  SPHERE optical system}. Additionally, identification between
Eqs.~\eqref{eq_reg} and \eqref{eq_regf2} yields the analytic value of
the hyperparameter $\mu_k$:
\begin{equation}\label{eq_hyper}
\mu_k=\frac{1}{\sigma_{\nabla\phi_k}^2},
\end{equation}
where $\sigma_{\nabla\phi_k}^2$ is defined as
$\sigma_{\nabla\phi_k}^2=\sigma_{\nabla_x\phi_k}^2+\sigma_{\nabla_y\phi_k}^2$,
with $\sigma_{\nabla_x\phi_k}^2$ and $\sigma_{\nabla_y\phi_k}^2$ the
variances of $\nabla(\phi_k)$ in directions $x$ and $y$, respectively.
The fact that the hyperparameter is given by Eq.~\eqref{eq_hyper} stems
from the assumption that the phase $\phi_k$ is statistically
homogeneous, and is whitened by the differentiation in Eq.
\eqref{eq_regf2} \cite{sauvage-p-06,mnras_11_sbongard}. One can notice
that $\sigma_{\nabla\phi_k}^2$ can be analytically computed from $S_{\phi_k}$
and the phase variance $\sigma_{\phi_k}^2$. Thus, this regularization does not
require any manual tuning.

\subsection{Coronagraphic image formation model}
\label{est_img}

To perform the minimization of criterion $J$ in Eq.~\eqref{eq-pb-inverse}, the
image formation model used by COFFEE (Eq.~\eqref{eq_im_model}) requires
the expression of a coronagraphic PSF $h_c$. Let $r$ be the pupil
plane position vector and $\gamma$ the focal plane position vector. the
entrance pupil function $P_u$ is such that:
\begin{equation}
P_u(r)=\Pi\left(\frac{2r}{D_u}\right)\Phi(r){ ,}
\end{equation}
with $\Pi$ the disk of unit radius, $D_u$ the entrance pupil diameter, and
$\Phi$ a known apodization function. The electric field in the entrance pupil
can be written as:
\begin{equation}
\Psi_A(r)=P_u(r)e^{j\phi_u(r)}\text{.}
\end{equation}
The electric field in the detector plane $\Psi_D$ is obtained by propagating
$\Psi_A$ through each plane of the coronagraphic imaging system: the signal is
first focused on the coronagraphic focal plane mask $\mathcal{M}$; then, the
electric field is propagated trough the Lyot Stop pupil $P_d(r)$
($P_d(r)=\Pi\left(2r/D_d\right)$ with $D_d$ the Lyot Stop pupil diameter).
The electric field in the detector plane $\Psi_D$ can thus be written as:
\begin{equation}\label{eq-field_coro}
  \Psi_d(\gamma)= \tf^{-1}\left\{ \tf\left[
        \tf^{-1}\left(\Psi_A(r)\right)\mathcal{M}\right] P_d(r)e^{j\phi_d(r)}\right\}{,}
\end{equation}
where $\tf^{-1}$ is the inverse Fourier transform operation. For the sake of
simplicity, spatial variables $r$ and $\gamma$ will be omitted in the following.

The coronagraphic PSF $h_c$ is the square modulus of $\Psi_D$:
\begin{equation}\label{eq-psf_coro}
  h_c= \left| \tf^{-1}\left\{ \tf\left[
        \tf^{-1}\left(\Psi_A\right)\mathcal{M}\right] P_de^{j\phi_d}\right\} \right|^2
\end{equation}

In Eq.~\eqref{eq-psf_coro}, $\mathcal{M}$ can easily be adapted to
represent any coronagraphic device, allowing COFFEE to be used with \review{a
  broad class of} high contrast imaging instruments.

\section{Performance evaluation}
\label{perf}
This Section presents the performance of the new extension of COFFEE presented in
Section \ref{est}. Section \ref{perf_param} gathers the different parameters
used for these simulations. In Section \ref{perf_noise}, the impact of the noise on
COFFEE's estimation is quantified, while the optimal phase diversity
$\phi_\inddiv$ to be used with COFFEE is studied in Section \ref{perf_div}.
\review{Section \ref{perf_psd} presents COFFEE's sensitivity to a difference
  between the prior assumed in the phase reconstruction and the true phase.}

\subsection{Parameters and criteria}
\label{perf_param}

\begin{figure}
\centering
\includegraphics[width = 0.25\linewidth]{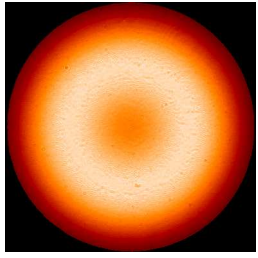}
\caption{Apodized Lyot coronagraph: apodization function used in this paper,
  computed from an experimental image recorded on SPHERE}
\label{fig_apod}
\end{figure}

Table \ref{table_est} gathers the parameters used for the simulations
presented in this section:

\begin{tiny}
\begin{table}
\centering
\begin{tabular}{m{7cm} m{6cm}}
\hline
image size & $64\times 64$ $\lovD$ ($128\times128$ pixels, Shannon-sampled)\\
Light spectrum & Monochromatic, wavelength $\lambda = 1589$ nm\\
Entrance pupil & $D_u = 64$ pixels\\
Lyot stop pupil & $D_d = D_u$\\
Aberration upstream of the coronagraph ($\phi_u$) & WFE$_u$ $=50$ nm RMS\\
Aberration downstream of the coronagraph ($\phi_d$) & WFE$_d$ $=20$ nm RMS\\
Coronagraph & Apodized Lyot Coronagraph (ALC), focal plane mask angular diameter $d=4.52\lovDt$\\
\hline
\end{tabular}
\caption{COFFEE: simulation parameters}
\label{table_est}
\end{table}
\end{tiny}
These parameters have been chosen so that the following simulations are
representative of the SPHERE instrument. The chosen coronagraph (ALC) is the
one designed for the considered wavelength on SPHERE, and the apodization
function used in the image formation model (Figure \ref{fig_apod}) is the one
designed for this coronagraph. \review{In this paper, we consider the case of
  a high-contrast imaging instrument calibration prior to the scientific
  observation, so a monochromatic source is considered. It is worth
  mentioning that COFFEE could easily be adapted to polychromatic images,
  although such a study is beyond the scope of this paper. Such an adaptation
  would require a modification of the image formation model (Eq.
  \eqref{eq_im_model}), in which a polychromatic coronagraphic PSF would be computed from
  several monochromatic PSF for different wavelengths.}
In order to properly model the ALC coronagraph, the coronagraphic PSF $h_c$
(Eq.~\eqref{eq-psf_coro}) is computed using the method developed by R.
Soummer \textit{et al.} \cite{SA_meth}. \review{This approach allows an accurate numerical
  representation of Lyot-style coronagraphs by accurately sampling the
  coronagraphic focal plane mask, which can hardly be done using the common
  Fast Fourier Transform (FFT) algorithm. Such an operation would indeed
  require the manipulation of very large arrays}.

For each simulation, coronagraphic images are computed from randomly generated
aberrations $\phi_u$ and $\phi_d$ using the imaging model presented in
Eqs~\eqref{eq_im_model} and \eqref{eq-psf_coro}. Using these two images,
COFFEE performs the phase estimation by minimizing criterion $J$ of
Eq.~\eqref{eq-pb-inverse}.

In order to quantify the reconstruction accuracy, we define the reconstruction
error $\epsilon_k$ as the RMS value of $\phi_k-\hat{\phi}_k$, where $k$ is
either $u$ (upstream) or $d$ (downstream), $\phi_k$ is the simulated
aberration and $\hat{\phi}_k$ its estimation made by COFFEE. In this Section, every
reconstruction error value is an average value, computed from ten independent
\review{randomly generated} phases \review{to make sure that the result
  obtained is independent of the phase realization}. The PSD $S_\phi$ of each
generated phase is such that $S_\phi \propto 1/\nu^2$.

\subsection{Noise propagation}
\label{perf_noise}

The ultimate limitation of the estimation performed by COFFEE lies in the
propagation of noise present in the images. As mentioned in Section
\ref{est_reg}, the use of a suitable regularization metric ensures the
smoothness of the phase, limiting the propagation of the noise from the images
to the estimated aberrations. In this Section, we first demonstrate that the
analytic value of the regularization metric hyperparameter (see Section
\ref{est_reg}) is the one that gives the smallest reconstruction error. We
consider here an incoming flux $\alpha=10^7$ photons and a read-out noise
(RON) of standard deviation $\sigma_\inddet=5\ \text{e}^-$. This value,
slightly higher than the expected RON on a SPHERE-like system, is chosen to
strengthen the impact of the hyperparameter value. Besides, photon noise is
added in the simulated coronagraphic images. The hyperparameter is here such
as $\mu_k=\chi/\sigma_{\nabla\phi_k}^2$, with $\chi \in [10^{-3} ; 10^{3}]$. A
reconstruction error value is computed for each value of $\chi$.

\begin{figure}
\centering
\includegraphics[width = 0.5\linewidth]{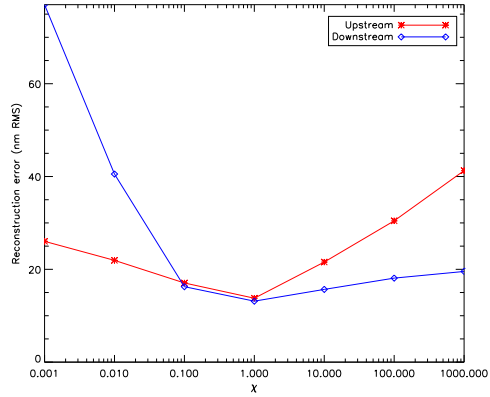}
\caption{Regularization metrics : reconstruction error upstream (red line) and
  downstream (blue line) of the coronagraph as functions of the hyperparameter
  value}
\label{fig_regul}
\end{figure}

Figure \ref{fig_regul} shows that the analytic value of the hyperparameter
($\chi = 1$) is the one that gives the smallest reconstruction error for both
upstream and downstream aberrations. When the regularization metrics are under
balanced ($\chi < 1$), the prior information is not ``strong'' enough in the
minimization to prevent noise propagation in the estimated aberrations. One
can notice here that without a regularization metric ($\chi=0$), the
reconstruction error would have been unacceptable. On the other hand, when the
regularization metrics are over balanced ($\chi > 1)$, their impact is too
strong, and prevents the estimation of the high frequency components of
$\phi_u$ and $\phi_d$. In the following, considering the result of this
simulation, all the estimations performed by COFFEE will be done using the
analytic value of the hyperparameter ($\chi = 1$).

\begin{figure}
\centering
\begin{tabular}{cc}
\includegraphics[width = 0.5\linewidth]{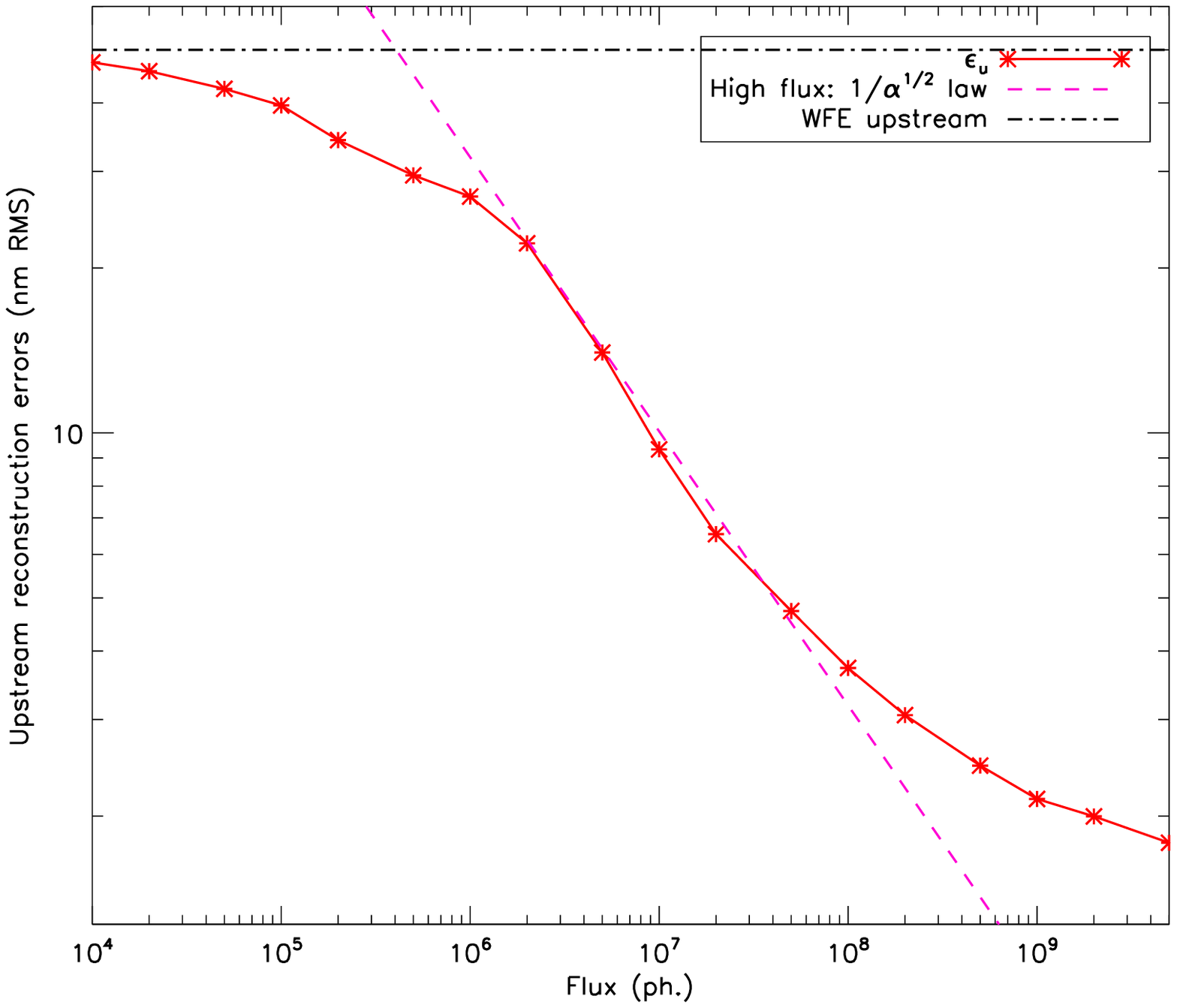}&
\includegraphics[width = 0.5\linewidth]{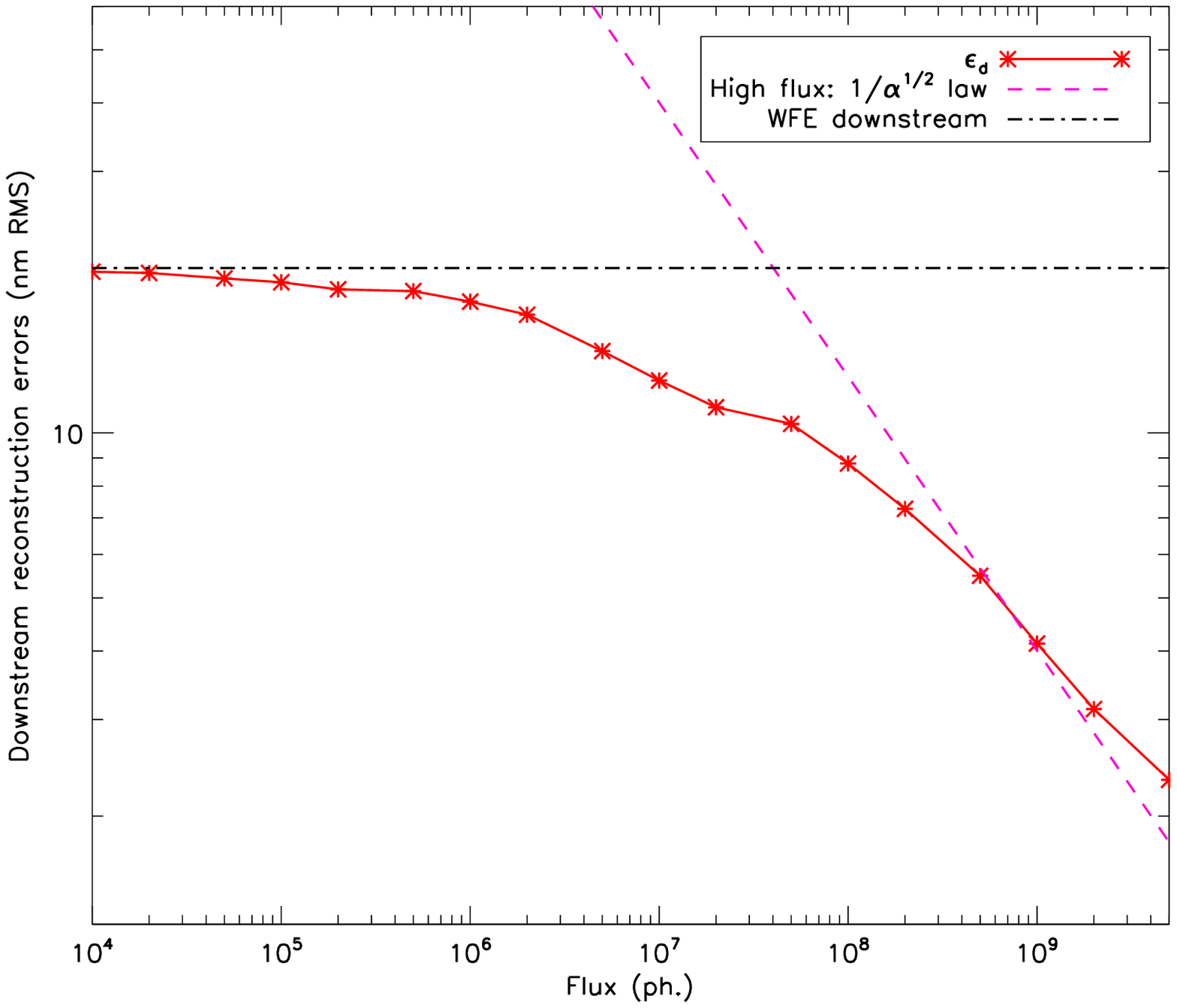}\\
\end{tabular}
\caption{Reconstruction error (solid red line) upstream (left) and downstream
  (right) as a function of the incoming flux $\alpha$. For comparison,
  $1/\sqrt{\alpha}$ (magenta dashed line) theoretical behaviour is plotted for
  photon noise only. The dotted-dashed black line represent the WFE value
  upstream and downstream of the coronagraph.}
\label{fig_flux}
\end{figure}

In Fig. \ref{fig_flux}, we present the evolution of the reconstruction
errors with respect to the total incoming flux. As previously, photon noise
and detector noise ($\sigma_\inddet=1\ \text{e}^-$) are added in the simulated
images used by COFFEE to perform the estimation.

In Fig. \ref{fig_flux}, one can see that in a low flux regime, both
reconstruction errors upstream and downstream reach a saturation level, which
correspond to the WFE value, rather than becoming arbitrarily large. Such a
behaviour originates in the regularization metric: in a low flux regime, the
speckles that originate in aberrations upstream of the coronagraph, which are
used by COFFEE to estimate the aberrations, can hardly be distinguished from
the noise. The less these speckles are visible in the images, the less the
corresponding aberrations can be estimated using COFFEE. The limit is reached
when no speckles are visible : in such a case, the estimated phase tend to
zero and the reconstruction error is equal to the WFE value. Notice that
without regularization metric, the reconstruction errors would have been much
stronger in a low flux regime, due to noise propagation (as presented in
Figure \ref{fig_regul}, in the case of an under-balanced hyperparameter).

When the flux increases, the reconstruction error evolution is proportional to
$1/\alpha$, which correspond to a photon noise limited regime. For very high
flux values ($\alpha > 10^9$ photons), the upstream
reconstruction error seems to reach another saturation level. We have shown
that this saturation, slightly above $1$ nm RMS can be attributed to numerical
difficulties in the minimization due to the very high dynamic range of the
noise variance in the criterion. If it were of practical interest, these
difficulties could be solved to reach even better accuracies.

\review{In this paper, we consider the case of a high-contrast imaging
  instrument calibration, performed off-line prior to the observation with a
  high SNR value. Thus, considering the noise propagation behaviour presented
  above, the following simulations will be performed with an incoming flux
  value $\alpha=10^9$ photons and a read-out noise (RON) of standard deviation
  $\sigma_\inddet=1\ \text{e}^-$.}

\subsection{Choice of a diversity phase}
\label{perf_div}

This section aims at studying the sensitivity of COFFEE to the diversity phase
$\phi_{\inddiv}$, which was until now a mix of defocus and astigmatism:
$\phi_\inddiv=a^\inddiv(Z_4 + Z_5)$ with $a^\inddiv=0.8\text{ rad RMS}$ ($202$
nm RMS at $\lambda=1589$ nm), introduced upstream of the coronagraph. This
choice has been made following Sauvage \textit{et al.} \cite{Sauvage-a-12}, who
demonstrated that for a perfect coronagraph model and low order aberrations,
such a diversity phase allowed a suitable criterion shape for the
minimization. Indeed, the use of this diversity phase instead of defocus alone
enlarges the global minimum, leading to an easier criterion minimization. In
this Section, we study the influence of the diversity phase on the
reconstruction accuracy more thoroughly and for realistic high order
aberrations and coronagraph.

\review{In classical phase diversity (no coronagraph), the optimal diversity
  phase depends on several parameters such as the signal to noise ratio (SNR),
  the level of the aberrations and their PSD\cite{Mugnier-l-06a}. A theoretical
  work, based on the computation of the Cramer-Rao lower bound (following Lee
  \textit{et al.}\cite{Lee-a-97}) could be performed to determine an optimal
  diversity phase; however, such a study would assume that there are no local
  minima in the criterion. Since we know that such minima appears in the
  criterion when the diversity phase amplitude is small, we adopt, in this
  section, a more practical approach to determine a suitable diversity phase
  for an aberration estimation with nanometric precision.}

We will consider different diversity phases: a diversity phase composed of
defocus alone, $\phi_\inddiv=a_\inddiv Z_4$, and a diversity phase composed of
a mix of defocus and astigmatism, $\phi_\inddiv=a^\inddiv(Z_4 + Z_5)$.
For each diversity phase, the evolution of the reconstruction errors with the
diversity phase amplitude $a_\inddiv$ value will be plotted for $3$ different
WFE$_u$ value upstream of the coronagraph.The parameters used in this simulation are
gathered in Table \ref{table_est}.

\begin{figure}
\centering
\begin{tabular}{cc}
\includegraphics[width = 0.5\linewidth]{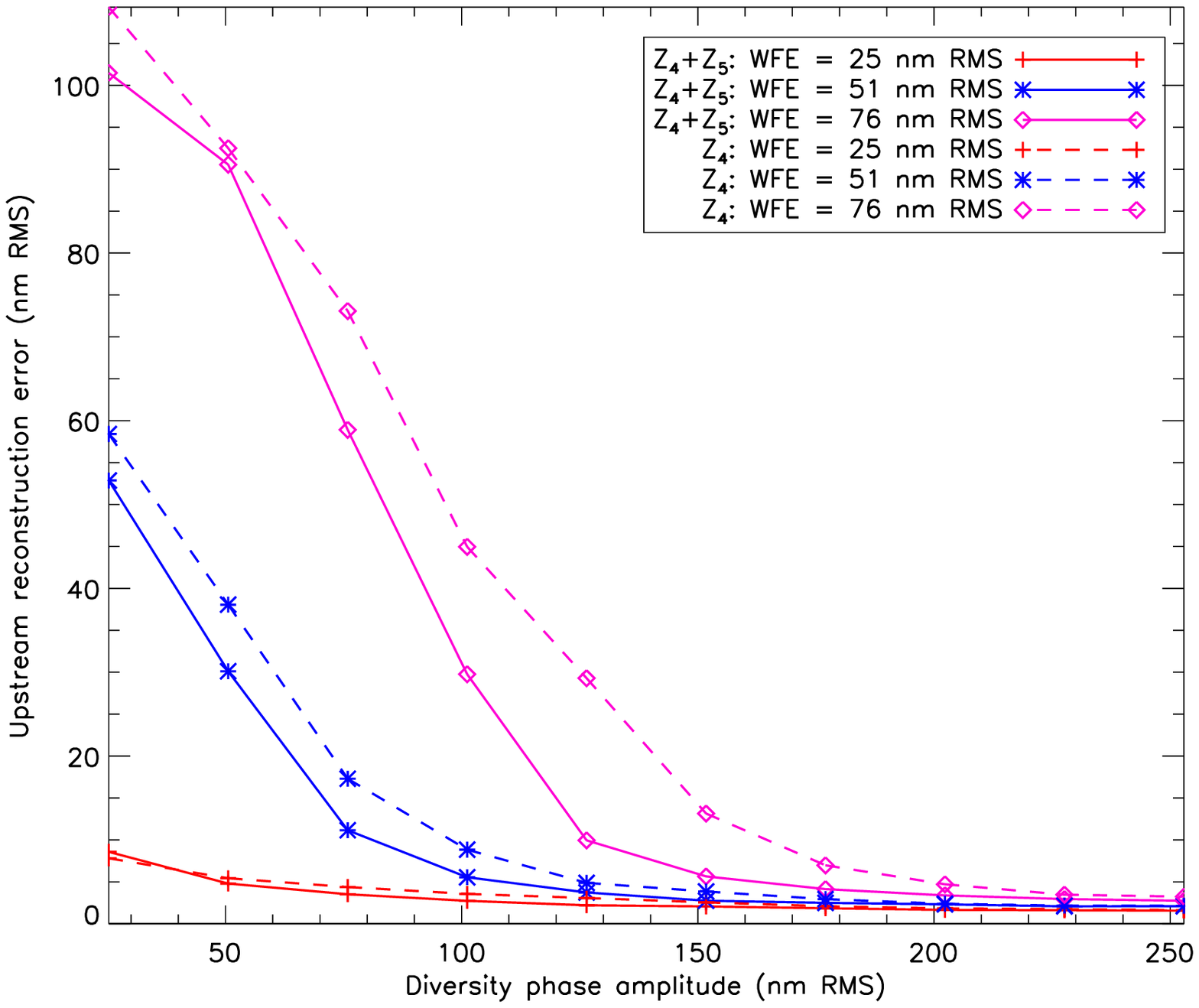}&
\includegraphics[width = 0.5\linewidth]{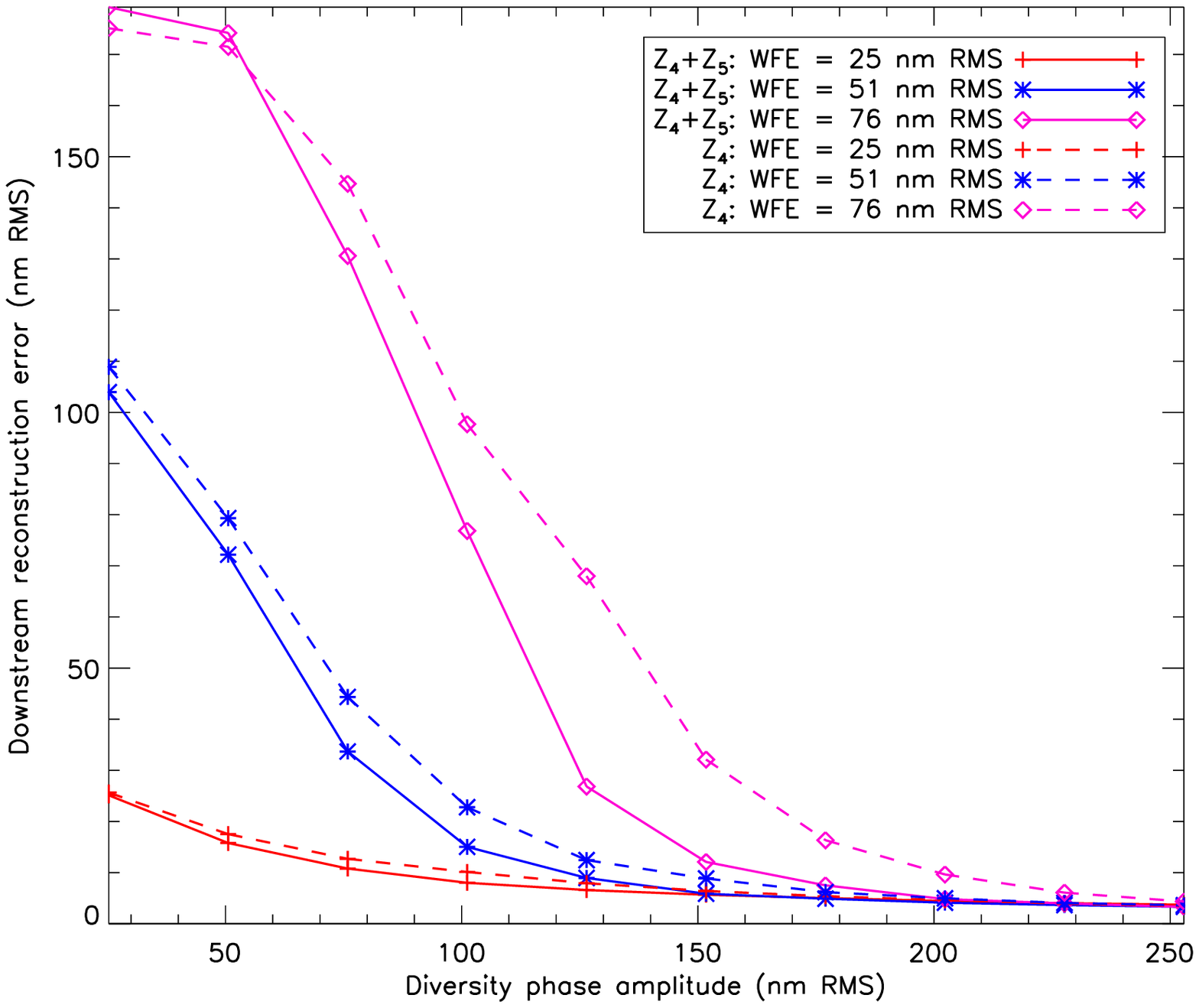}\\
\end{tabular}
\caption{Reconstruction errors upstream (left) and downstream (right) of the
  coronagraph as functions of the amplitude $a^\inddiv$ of a diversity phase
  composed of defocus only ($Z_4$, dashed lines), or composed of a mix of
  defocus and astigmatism ($Z_4+Z_5$, solid lines).}
\label{fig_divopt_dda}
\end{figure}
 
Figure \ref{fig_divopt_dda} shows the evolution of both reconstruction errors
upstream and downstream of the coronagraph with respect to the amplitude
$a^\inddiv$. Here, the reconstruction error is due to noise and to local
minima, which are gradually removed when the diversity phase amplitude
increases, leading to an improvement of the estimation accuracy. When the
diversity phase amplitude $a^\inddiv$ is high enough, all local minima are
removed. Then, the reconstruction error reaches a saturation level which
correspond to the level of noise in the images.

One can notice that with a diversity composed of defocus and astigmatism, the
reconstruction error decreases faster than in the case of a diversity composed
of a defocus alone. In order to have $\epsilon_u < 5$ nm RMS, $a_{div}$ must
be greater than $2\times$WFE$_u$ when the diversity is composed of defocus and
astigmatism, and greater than $2.5\times$WFE$_u$ when it is composed of defocus alone.

This result confirms what has been shown in a simple case
with a perfect coronagraph model \cite{Sauvage-a-12}: a diversity phase
composed of a mix of defocus and astigmatism enlarges the global minimum and
pushes away local minima, making criterion $J$ (Eq.~\eqref{eq-pb-inverse})
easier to minimize, and thus allowing a more accurate estimation of the
aberrations both upstream and downstream of the coronagraph.

\subsection{\review{Sensitivity to the \textit{a priori} assumption accuracy}}
\label{perf_psd}

\review{As mentioned in Section \ref{est_reg}, the \textit{a priori}
  assumption about the aberration PSD considered by COFFEE is derived from
  \cite{Dohlen-p-11}. However, this assumption will not be exact for all
  optical surface; for instance, the PSD that correspond to the polishing
  error of the Very Large Telescope primary mirrors follow a $1/\nu^3$ power
  law, as demonstrated by Bord\'e and Traub\cite{speckle_nulling}.}

\review{In this section, the sensitivity of COFFEE to the validity of the
  assumed PSD is evaluated: using parameters listed
  in Table \ref{table_est}, coronagraphic images are computed with
  aberrations generated with three different power laws: $1/\nu$, $1/\nu^2$
  and $1/\nu^3$. Then, using these simulated images, COFFEE perform the
  aberration estimation assuming a PSD following a $1/\nu^2$
  power law (as described in Section \ref{est_reg}).}

\begin{tiny}
\begin{table}
\centering
\begin{tabular}{|c|c|c|}
\hline
Aberration's PSD & $\epsilon_u$ & $\epsilon_d$ \\
\hline
$S_{\phi_k} \propto 1/\nu$  & $3.08$ nm RMS & $5.85$ nm RMS \\
$S_{\phi_k} \propto 1/\nu^2$  & $2.28$ nm RMS & $4.09$ nm RMS \\
$S_{\phi_k} \propto 1/\nu^3$  & $1.54$ nm RMS & $3.40$ nm RMS \\
\hline
\end{tabular}
\caption{\review{Comparison of COFFEE's estimation accuracy when the \textit{a priori}
  knowledge on the aberration's PSD is not perfectly accurate. As previously,
  $k$ stands for $u$ (upstream) or $d$ (downstream).}}
\label{table_est_psd}
\end{table}
\end{tiny}

\review{Table \ref{table_est_psd} shows the results of this evaluation. One
  can see that COFFEE is not very sensitive to the \textit{a priori} accuracy,
  since the reconstruction errors remain indeed small when the PSD assumed by
  COFFEE does not perfectly match the actual PSD.}

\review{When $S_{\phi_k}$ follows a $1/\nu$ power law, we note a slight
  increase in the reconstruction error. We have checked that, as expected,
  this increase is due to very high frequencies in the phase which are
  over-regularized and thus not reconstructed. In the conditions of our
  simulations, these very high frequencies are actually beyond the spatial
  frequencies that the DM is able to control, so that in closed loop this HF
  error on the reconstructed phase would have no impact.}

\review{We note that when $S_{\phi_k}$ follow a $1/\nu^3$ power law,
  reconstruction errors are lower than when the correct $1/\nu^2$ power law is
  considered. Again, the reconstruction errors originate mostly in the
  estimation of the high frequency aberrations, which give birth to lower
  energy speckle than the low frequency aberrations. When the PSD
  follows a $1/\nu^3$ power law, the quantity of high frequencies decreases in
  the aberrations to estimate, leading to an improved reconstruction error.}

\section{Circumventing calibration errors of the diversity phase: the myopic approach}
\label{myop}

The value of the diversity phase $\phi_\inddiv$ is one of the few inputs
COFFEE needs to perform the phase estimation. Thus, an imprecise calibration
of $\phi_\inddiv$ will lead to an error on the estimated aberration. In the
case of classical phase diversity, this error is the one that drives the total
error budget \cite{Blanc-a-03a}. In \cite{Paul-a-13}, we demonstrated that an
error $\epsilon_\inddiv$ on the knowledge of $\phi_\inddiv$ was leading to a
reconstruction error of about $\epsilon_\inddiv/2$ on both upstream and
downstream estimated aberrations. Now that both aliasing and model errors have
been tackled in COFFEE, the diversity calibration error would be the most
important one in the error budget.

The most convenient way to introduce the diversity phase $\phi_\inddiv$ on the
instrument is to modify the reference slopes of the AO loop to introduce a
calibrated aberration, as described in \cite{Paul-a-13}. The accuracy of such
a process is thus limited by the DM's ability to achieve a
given shape, leading to an error $\epsilon_\inddiv^\inddm$ on the phase
diversity $\phi_\inddiv^\inddm$ actually introduced. As this error will always
be present on an AO system (thermal evolution of the DM, ageing of influence
functions, inability to re-calibrate them regularly), we have adapted COFFEE
to make it able to perform a joint estimation of an error on $\phi_\inddiv$.
This approach, called hereafter ``myopic estimation'', consist in a slight
modification of the criterion $J$ to be minimized
(Eq.~\eqref{eq-pb-inverse}), in which an additional unknown parameter
$\phi_\epsilon$ (called hereafter diversity error phase) is introduced:
\begin{equation}\label{eq-pb-inverse_sec}
\begin{aligned}
J(\alpha, \beta, \phi_u, \phi_d, \phi_\epsilon)&=\frac{1}{2} \norm{\frac{i_c^{\text{foc}} 
- (\alpha_\indfoc h_\inddet\star h_c(\phi_u,\phi_d)+\beta_\indfoc)}{\sigma_n^{\text{foc}}}}^2 \\
&+\frac{1}{2}\norm{\frac{i_c^{\text{div}} 
- (\alpha_\inddiv h_\inddet\star h_c(\phi_u+\phi_\inddiv^\indcal+\phi_\epsilon,\phi_d)+\beta_\inddiv)}{\sigma_n^{\text{div}}}}^2\\
&+\mathcal{R}(\phi_u) + \mathcal{R}(\phi_d) + \mathcal{R}(\phi_\epsilon)\text{,}
\end{aligned}
\end{equation}
where $\phi_\inddiv^\indcal$ is the calibrated diversity phase:
$\phi_\inddiv=\phi_\inddiv^\indcal+\phi_\epsilon$.
$\mathcal{R}(\phi_\epsilon)$ is an optional regularization metric designed to
enforce our knowledge that $\phi_\epsilon$ should be small and smooth. Using
the gradient $\grad{J}{\phi_\epsilon}$ (whose computation is trivial knowing
$\grad{J}{\phi_u}$), COFFEE is able to perform a joint estimation of
$\phi_\epsilon$ along with the previously estimated parameters $\alpha, \beta,
\phi_u$ and $\phi_d$.

In this paper, the estimated phases $\phi_u$ and $\phi_d$ are expanded on a
pixel basis, which allows the estimation of high-order aberrations. However,
since the diversity phase is composed of low order aberration, one can expect
that the error $\phi_\epsilon$ will be mainly composed of low order
aberrations. Thus, in order to optimize COFFEE's performance, the estimation
of $\phi_\epsilon$ has been implemented in two different ways: 
\itemdiam{
\item $\phi_\epsilon$ can be expanded on a pixel basis if high-order patterns
  (such as dead actuators) are expected in the diversity error phase. In this
  case, the regularization metric $\mathcal{R}(\phi_\epsilon)$ used will have
  the same expression as $\mathcal{R}(\phi_u)$, presented in Section \ref{est_reg}.\\
\item If only low order aberrations are expected in $\phi_\epsilon$, it can be
  expanded on a truncated Zernike basis, composed of a few Zernike modes
  (typically up to $5$ if $\phi_\inddiv$ is composed of defocus ($Z_4$) and
  astigmatism ($Z_5$)). With this basis, which reduces the number of
  parameters to be estimated, no regularization metric dedicated to
  $\phi_\epsilon$ is needed. }

We validate this myopic approach by a realistic simulation: using parameters
gathered in Table \ref{table_est}, we simulate coronagraphic images, and
consider that the diversity phase used for the simulation is not perfectly
known in the estimation stage. The coronagraphic simulated diversity image is
computed with a diversity $\phi_\inddiv=\phi_\inddiv^\indcal+\phi_\epsilon$,
where $\phi_\epsilon=a_\epsilon(Z_4+Z_5)$, considering that the amplitude of
the phase diversity is not perfectly known. In this simulation,
$a_\epsilon=0.04$ rad RMS ($10$ nm RMS).

COFFEE's phase reconstruction is then performed using both simulated images
considering that the diversity phase is equal to $\phi_\inddiv^\indcal$, first without
the myopic estimation, then with an estimation of $\phi_\epsilon$ expanded on
pixel basis and on a truncated Zernike basis composed of $4$ modes: tip, tilt
(which allows an estimation of a differential tip-tilt
between the two images), defocus and astigmatism.\\

\begin{tiny}
\begin{table}
\centering
\begin{tabular}{|c|c|c|}
\hline
& $\epsilon_u$ & $\epsilon_d$ \\
\hline
$\phi_\epsilon:$ no estimation & $6.92$ nm RMS & $8.51$ nm RMS \\
$\phi_\epsilon:$ estimation on a pixel-wise map & $2.55$ nm RMS & $5.45$ nm RMS \\
$\phi_\epsilon:$ estimation on $4$ Zernike modes & $2.64$ nm RMS & $4.16$ nm RMS \\
\hline
\end{tabular}
\caption{Comparison of COFFEE estimation accuracy with and without myopic
  estimation of $\phi_\epsilon$ when the diversity phase is not perfectly known.}
\label{table_est_err}
\end{table}
\end{tiny}

The reconstruction errors corresponding to each reconstruction are gathered in
Table \ref{table_est_err}. One can clearly see here that the myopic estimation
approach significantly improves the reconstruction accuracy, roughly by a
factor $2$, allowing an optimal use of COFFEE even when the diversity phase
$\phi_\inddiv$ is not perfectly known.

\review{The usefulness of this myopic approach will be further illustrated in Section
  \ref{comp_dh}, where we show that the estimation of an error on the
  diversity phase allows us to improve the contrast on the detector plane in a
  compensation process.}

\section{Closed loop quasi-static aberration compensation using COFFEE}
\label{comp}

In this Section, we present COFFEE's performance in a closed loop compensation
process. We consider here the case of the calibration of a SPHERE-like
instrument. Two coronagraphic images are simulated using randomly generated
aberrations, whose PSD follows a $1/\nu^2$ law. Then, using these two
simulated coronagraphic images, the aberrations upstream $\phi_u$ and $\phi_d$
are estimated using COFFEE.

Once the phase reconstruction is performed, we modify the DM actuator voltages
to compensate for the estimated aberrations and thus optimize the contrast in
the detector plane. This compensation is performed using two different
techniques: in Section \ref{comp_conjphi}, we use the conventional phase
conjugation. In Section \ref{comp_dh}, we minimize the energy in a chosen area
in order to optimize the contrast in the selected region of the detector
plane. To perform such a compensation, we have developed a method dedicated to
high-contrast imaging instruments that does not rely on any small aberration
approximation. These simulations are performed using the parameters gathered
in Tables \ref{table_est} and \ref{table_comp}.

\begin{tiny}
\begin{table}
\centering
\begin{tabular}{m{7cm} m{6cm}}
\hline
Lyot stop pupil & $D_d = 0.96D_u$\\
Incoming flux & $10^9$ photons\\
Detector noise & $\sigma_{\inddet}=1\ \text{e}^-$\\
Deformable mirror (DM) & $41\times 41$ actuators, Gaussian-shaped influence functions\\
\hline
\end{tabular}
\caption{COFFEE: parameters used for the compensation simulations of Section \ref{comp}}
\label{table_comp}
\end{table}
\end{tiny}

In order to perform a realistic simulation, we consider here that an error
$\phi_\epsilon$ is made on the diversity phase: the coronagraphic diversity image
is computed with a diversity phase
$\phi_\inddiv^{\text{sim}}=\phi_\inddiv+\phi_\epsilon$. In this section, we
consider that $\phi_\epsilon=a_\epsilon(Z_4+Z_5)$, with
$a_\epsilon=0.04$ rad RMS ($10$ nm RMS). \\
COFFEE estimates the aberrations considering that the calibrated diversity phase is equal
to $\phi_\inddiv$, and jointly searches for the diversity phase error  $\phi_\epsilon$, as
described in Section \ref{myop}.

\subsection{Phase conjugation}
\label{comp_conjphi}

Conventional phase conjugation aims at compensating for the aberrations
upstream of the coronagraph $\phi_u$ in an iterative process. After its
criterion minimization, COFFEE gives an estimation $\hat{\phi}_u$. The
aberrations upstream of the coronagraph at the iteration $i+1$ are thus given by:
\begin{equation}
\phi_u^{i+1}=\phi_u^{i}-g\hat{\phi}_u^{\inddm}\text{,}
\end{equation}
where $g$ is the gain of the iterative process ($g=0.5$ in this simulation)
and $\hat{\phi}_u^{\inddm}$ is the aberration introduced by the DM in the
entrance pupil plane. Such an aberration corresponds to the best
representation of $\hat{\phi}_u$ achievable by the deformable mirror. Let $F$
be the DM's influence matrix and $T$ its generalized inverse. The aberration
introduced by the DM can be computed as follow:
\begin{equation}
\hat{\phi}_u^{\inddm}=FT\hat{\phi}_u.
\end{equation}
\begin{figure}
\centering
\begin{tabular}{cc}
\includegraphics[width = 0.2\linewidth]{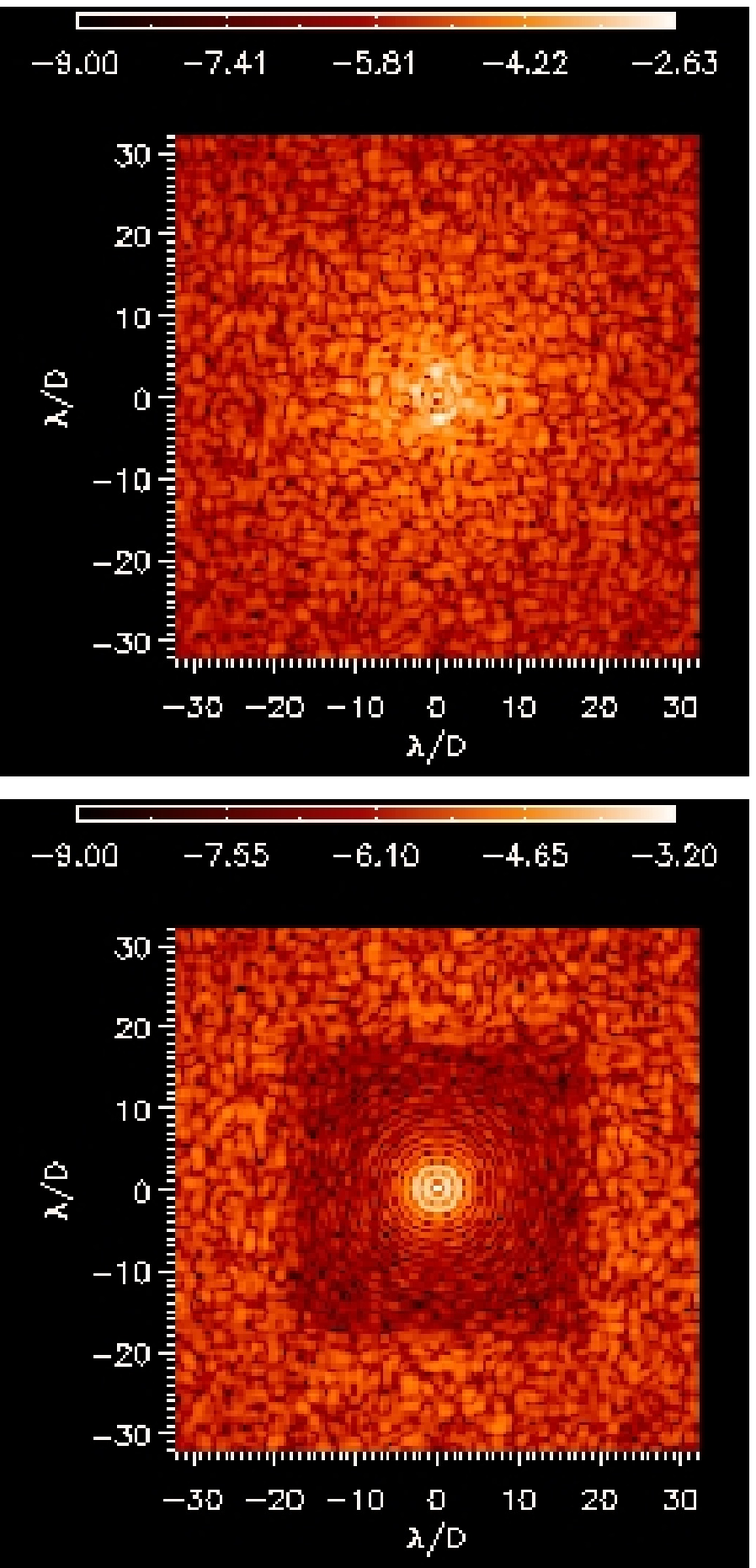}&
\includegraphics[width = 0.5\linewidth]{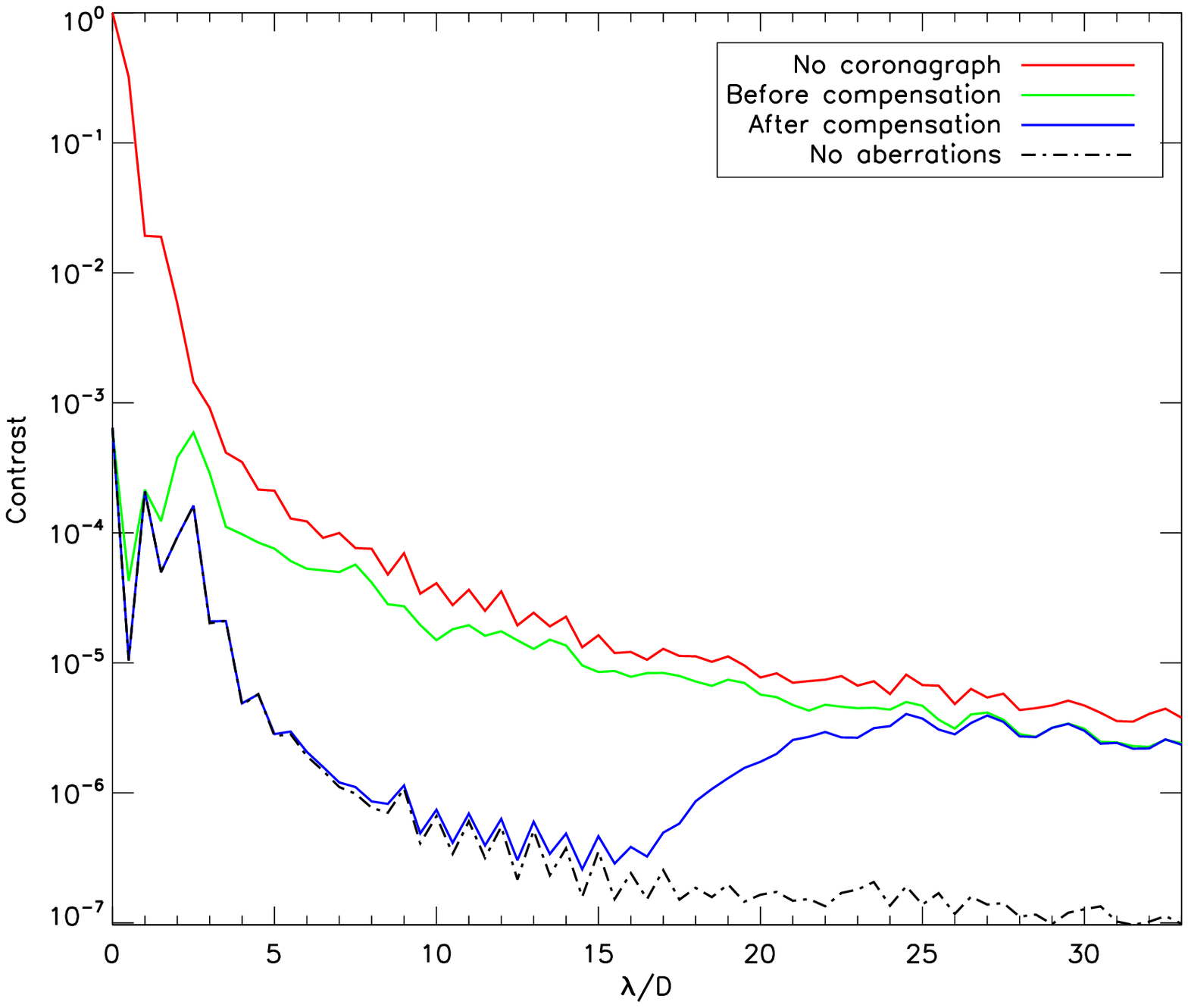}\\
\end{tabular}
\caption{COFFEE : simulation of aberration compensation using classical phase
  conjugation. Left : coronagraphic images before (top) and after five
  iterations of the closed loop process (bottom, logarithmic scale for both
  images). Right: normalized average raw contrast computed from the images
  without coronagraph (solid red line), with coronagraph before (solid green
  line) and after compensation (solid blue line). For comparison, the contrast
  computed from a coronagraphic image computed without any aberrations is
  plotted as well (dashed black line).}
\label{fig_conjphi}
\end{figure}

Figure \ref{fig_conjphi} presents the result of the simulation of a
compensation performed by phase conjugation \review{after $5$ iterations of
  the loop (the average computation time for one loop iteration is $2$
  minutes)}. On this figure, one can see that the aberration compensation
performed using COFFEE estimation allows a significant improvement in the
coronagraphic images (figure \ref{fig_conjphi}, left). In particular, the
average contrast plot (figure \ref{fig_conjphi}, right) shows that after
compensation, the performance in the area controlled by the DM ($\pm
20\lovDt$) is very close to the one that would be obtained from a
coronagraphic image computed without aberrations.

We note that the level of contrast reached after compensation presented in
Fig. \ref{fig_conjphi} exceeds the SPHERE instrument specification for
off-line calibration, which relies on a phase estimation based on classical
phase diversity (no coronagraph). In particular, classical phase diversity is
not able to estimate high-order frequencies, and thus will not compensate for
speckles located beyond $8\lovDt$. The use of COFFEE, which allows a
compensation in the whole area controlled by the DM in the detector plane (as
showed in Fig. \ref{fig_conjphi}), could thus improve the SPHERE instrument
performance.

\subsection{Creation of a Dark Hole on the detector}
\label{comp_dh}

Speckle nulling iterative techniques aim at minimizing the energy in a chosen
area of the detector in order to facilitate exoplanet detection in this area,
called a ``Dark Hole'' (DH). To create this DH, the methods developed until
now, which rely on a small aberration approximation, minimize the energy
during an iterative process \cite{speckle_nulling,efc}. This process
requires several iterations and is based on the knowledge of an interaction
matrix between the detector plane and the DM pupil plane to create the Dark
Hole on the detector.

Here, we propose a new method to minimize the energy that does not rely on any
small aberration assumption\review{, allowing us to deal with high amplitude
  phase aberrations such as the ones created by the DM dead actuators (which
  are currently part of the SPHERE instrument's limitations). Besides, this
  method takes into account both upstream and downstream aberrations}. This
compensation method, coupled with COFFEE would be particularly adapted to the
calibration of a high-contrast imaging instrument.

Let us define the energy in the focal plane $\En_\inddh$ in the DH as:
\begin{equation}\label{En_dh}
\En_\inddh=  \alpha\sum_{m,n\in \text{DH}}\module{\Psi_\inddh(m,n)}^2
\end{equation}
With $m,n$ the pixel position in the DH (for the sake of simplicity, these
variables will be omitted in the following). $\alpha$ is the incoming flux and
$\Psi_\inddh$ the electric field in the DH area which, using notations of
Section \ref{est_img}, is given by:
\begin{equation}\label{eq-field_dh}
  \Psi_\inddh(\volt)= \tf^{-1}\left\{ \tf\left[
        \tf^{-1}\left(P_ue^{j(\phi_u+\psi(\volt))}\right)\mathcal{M}\right] P_de^{j\phi_d}\right\}.
\end{equation}
$\psi$ represents the aberration introduced by the DM: $\psi(\volt)=F\volt$,
with $\volt$ the set of voltages applied to the DM actuators. Thus, creating a
DH on the detector means finding the set of voltages $\volt_\inddh$
that minimize the energy $\En_\inddh(\volt)$, knowing the relation between the
entrance pupil plane and the detector plane of the high contrast imaging
system.

Here, we minimize $\En_\inddh(\volt)$ by means of the same optimization method
as the one which was used by COFFEE to perform its estimation, which is a
limited memory variable metric (BFGS) method \cite{numerical_recipes,
  Thiebaut-p-02}. Such an operation gives us the voltage $\volt_\inddh$, and
thus the aberration $\psi(\volt_\inddh)$ to introduce using the DM to create
the DH. \review{Unlike others energy minimization methods, the one we propose
  here does not require an interaction matrix calibration between the detector
  plane and the DM pupil plane. In particular, using COFFEE, we are able to
  estimate the aberrations downstream of the coronagraph, and among them the
  coronagraphic PSF position on the detector, i.e. downstream tip-tilt. This
  estimation is then taken into account in our compensation method (Eqs
  \eqref{En_dh} and \eqref{eq-field_dh}). Thus, variation in the
  aberrations downstream of the coronagraph will not require any particular
  action during the compensation process, since these variations will be
  estimated by COFFEE along with the aberrations upstream of the coronagraph.}

As mentioned previously, COFFEE does not estimate amplitude aberrations.
However, the compensation method described in this section can easily be
adapted to take into account amplitude aberrations upstream of the coronagraph
$\xi$, by modifying Eq. \eqref{eq-field_dh}:
\begin{equation}
  \Psi_\inddh(\volt)= \tf^{-1}\left\{ \tf\left[
        \tf^{-1}\left(P_ue^{j(\phi_u+\psi(\volt))+\xi}\right)\mathcal{M}\right] P_de^{j\phi_d}\right\}
\end{equation}
With $\xi$ the amplitude aberrations. Thus, provided these aberrations are
known, amplitude aberrations does not limit the performance of the
compensation method presented in this Section.\\

We now validate this new energy minimization method by simulation. It requires
the knowledge of the aberrations upstream $\phi_u$ and downstream $\phi_d$ of
the coronagraph, and of the incoming flux $\alpha$. In this section, we use
COFFEE to perform the estimation of these parameters, and then minimize
the energy in order to create a DH in the detector plane. The energy
$\En_\inddh$ is minimized between $5\lovD$ and $20\lovD$ to create the Dark
Hole in the right part of the focal plane. \review{As in the phase conjugation
  case (Section \ref{comp_conjphi}), such a compensation is performed in an
  iterative process, where the aberrations upstream of the coronagraph at the
  iteration $i+1$ are given by:
\begin{equation}
\phi_u^{i+1}=\phi_u^{i}+gF\volt_\inddh\text{.}
\end{equation}
As previously, we consider $g=0,5$ in this simulation.}

\begin{figure}
\centering
\begin{tabular}{cc}
\includegraphics[width = 0.2\linewidth]{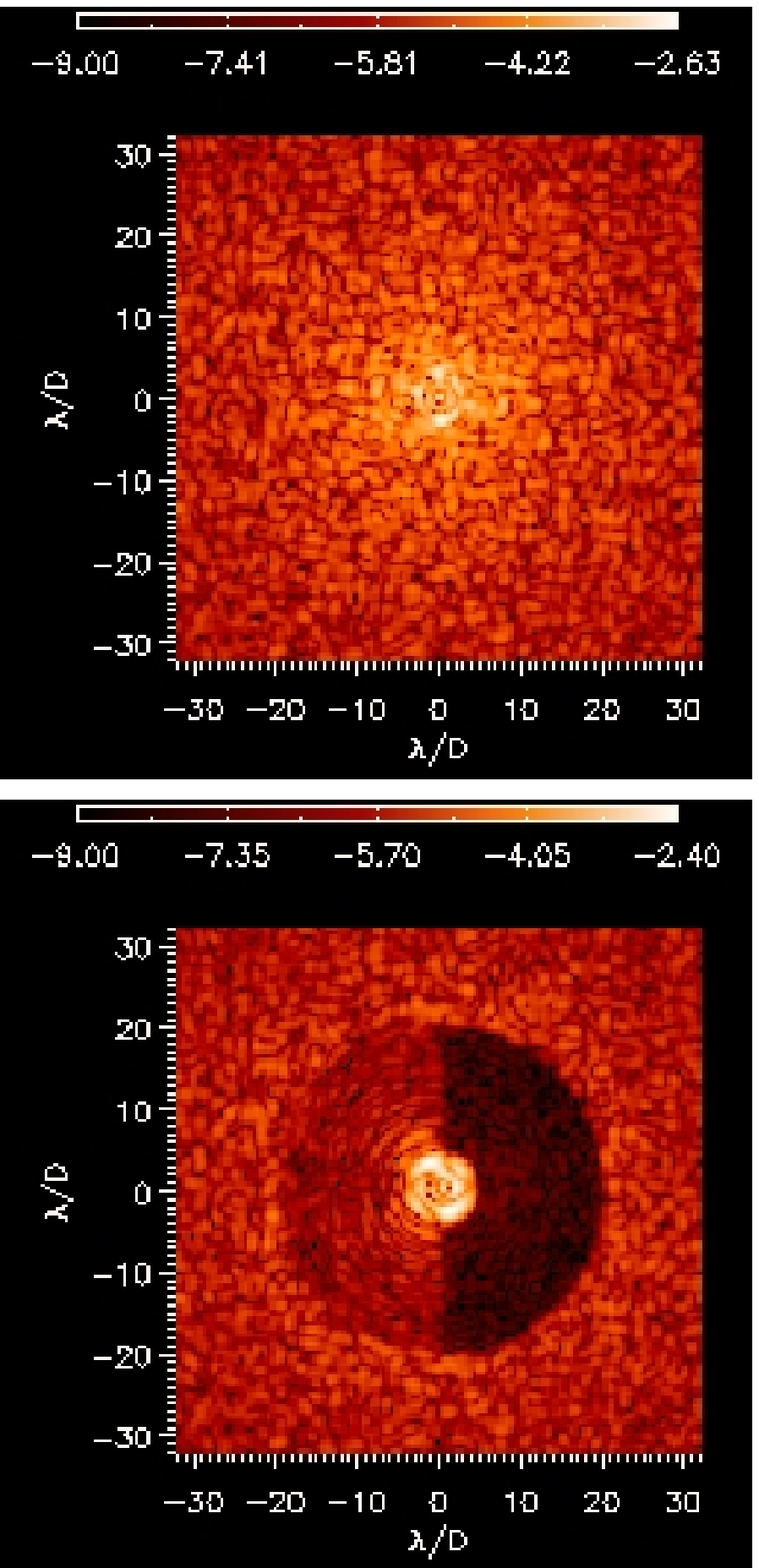}&
\includegraphics[width = 0.5\linewidth]{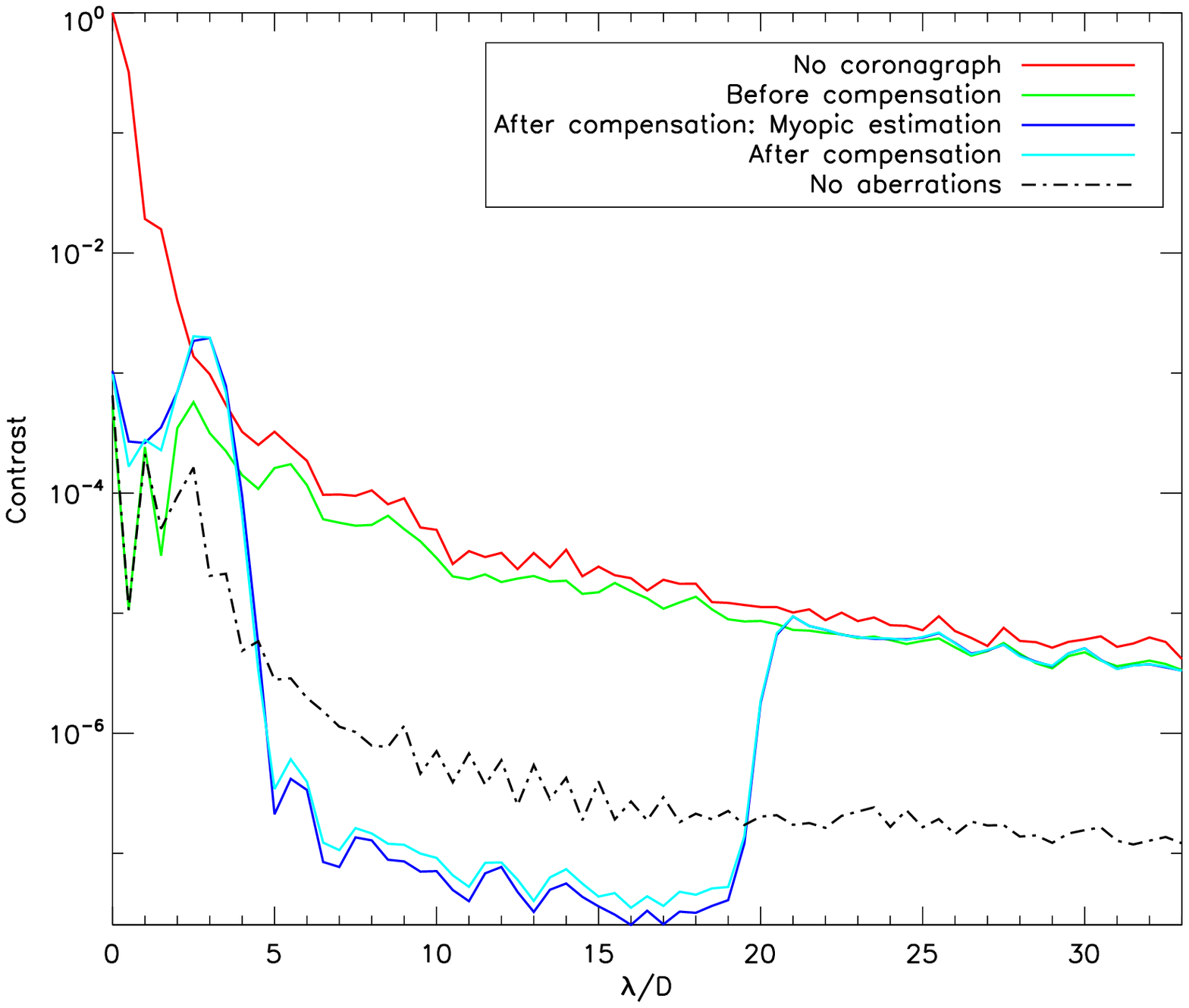}\\
\end{tabular}
\caption{COFFEE : simulation of aberration compensation by minimizing the
  energy in a chosen area (between $5\lovD$ and $20\lovD$ in the right part of
  the focal plane). Left : coronagraphic images before (top) and after five
  iterations of the closed loop process (bottom, logarithmic scale for both
  images). Right: normalized average raw contrast computed in the dark hole
  area from the images without coronagraph (solid red line), with coronagraph
  before (solid green line) and after compensation. After
  minimization, the average contrast in the Dark Hole is $3.1\
  10^{-8}$ \review{when the myopic approach is used (solid blue line), and
    $6.6\ 10^{-8}$ otherwise (solid cyan line)}. For comparison, the contrast computed from a
    coronagraphic image computed without any aberrations is plotted as well
    (dashed black line).}
\label{fig_dh}
\end{figure}

The result of this compensation \review{after $5$ iterations of the loop} is
presented in Fig. \ref{fig_dh} \review{(the average computation time for one
  loop iteration is $3$ minutes)}. In the targeted area (between $5\lovD$ and
$20\lovD$), this method allows a significant improvement: when the
compensation is performed by conventional phase conjugation (Figure
\ref{fig_conjphi}), the average contrast in the same area is $7.2\ 10^{-7}$.
The use of our new compensation method allow a contrast improvement by a
factor $10$ after compensation (Figure \ref{fig_dh}, bottom left).
\review{Besides, the interest of the myopic approach is illustrated in Fig.
  \ref{fig_dh} : when the closed loop process is performed without estimation
  of an error on the diversity phase, the performance decreases by a factor
  $2$ (solid cyan line).}

If a very high contrast level is required, this compensation technique can
thus be used instead of conventional phase conjugation to calibrate the
instrument. Besides, in order to push down the dark hole floor, it is possible
to narrow the energy minimization area, as mentioned by Bord\'e \textit{et al.}
\cite{speckle_nulling}.

\section{Conclusion}
\label{ccl}

In this paper, an extended version of our coronagraphic phase diversity,
nicknamed COFFEE, has been presented. The use of a regularized pixel basis in
the estimation allows COFFEE to estimate high order aberrations with
nanometric precision (Section \ref{est}). Besides, thanks to a modification of
the coronagraphic PSF used in the imaging model, COFFEE is no longer limited
to a particular coronagraphic device. COFFEE's performance has been studied
and discussed in Section \ref{perf}, while Section \ref{myop} has described a
so-called myopic extension of COFFEE, which consists in a joint estimation of
an error on the diversity phase in order to improve COFFEE's accuracy in a
real system, where the diversity phase is not perfectly known. Lastly, in
Section \ref{comp}, the achievable contrast optimization on a SPHERE-like
system using COFFEE in a compensation process has been studied using realistic
simulations. In the latter section, we have presented a new compensation
method which minimizes the energy in a chosen area of the detector through a
non-linear minimization, in order to reach higher level of contrast than those
that can be obtained using phase conjugation.\\

The experimental validation of this high-order and myopic version of COFFEE is
ongoing, and aims at demonstrating the ability of COFFEE to estimate both low
and high order aberrations, and to compensate for them. Several perspectives
are currently considered for this work. With an adaptation of the
coronagraphic imaging model, COFFEE can be extended to work on ground-based,
long exposure images with residual turbulence induced aberrations. Another
perspective lies in optimization of the computation time required for the
aberration estimation, possibly following I. Moc\oe{}ur \textit{et al.}
\cite{Mocoeur-a-09b}. These two improvements will allow COFFEE to work
on-line, in closed loop during the scientific exposure. A further perspective
is to extend COFFEE to the estimation of amplitude aberrations, mandatory to
reach the very high levels of contrast required for exo-earth imaging,
in order to create a dark hole area on the
detector using the method presented in this paper.\\

\section*{Acknowledgments}

The authors would like to thank several key players of the SPHERE instrument,
namely Thierry Fusco, David Mouillet, Jean-Luc Beuzit and Marc Ferrari for
simulating discussions and support, and the R\'egion Provence-Alpes-C\^ote d'Azur for
partial financial support of B. Paul's scholarship. This work was partly
funded by the European Commission under FP7 Grant Agreement No. 312430 Optical
Infrared Coordination Network for Astronomy.

\appendix

\section{\review{Gradients expression}}
\label{app_grad}

\review{The numerical minimization of criterion $J$ (Eq. \eqref{eq-pb-inverse})
requires the analytic expression of gradients $\grad{J}{\phi_u}$,
$\grad{J}{\phi_d}$, $\grad{J}{\alpha}$ and $\grad{J}{\beta}$ to estimate the
aberrations upstream $\phi_u$ and downstream $\phi_d$ of the coronagraph, as
well as the incoming flux $\alpha$ and the residual background $\beta$. Let us
rewrite here the expression of criterion $J$:}
\begin{equation}\label{eq-pb-inverse_ap}
\begin{aligned}
J(\alpha, \beta, \phi_u, \phi_d)&=\frac{1}{2} \norm{\frac{i_c^{\text{foc}} 
- (\alpha_\indfoc h_\inddet\star h_c^{\text{foc}}
+\beta_\indfoc)}{\sigma_n^{\text{foc}}}}^2  +\frac{1}{2}\norm{\frac{i_c^{\text{div}} 
- (\alpha_\inddiv h_\inddet\star h_c^{\text{div}} +\beta_\inddiv)}{\sigma_n^{\text{div}}}}^2\\
&+\reg(\phi_u) + \reg(\phi_d)\\
&= J^{\text{foc}}+J^{\text{div}}+\reg(\phi_u) + \reg(\phi_d)
\end{aligned}
\end{equation}
\review{The expressions of $\grad{J}{\alpha}$ and $\grad{J}{\beta}$ can be found in
\cite{Paul-a-13}. The calculation of gradients $\grad{J}{\phi_u}$ and
$\grad{J}{\phi_d}$ is performed following what have been done in
\cite{Paul-a-13}: we derive $J^{\text{foc}}$, and then deduce the gradients' expressions of
$J^{\text{div}}$ using a trivial substitution. The notations used here are the
ones introduced in Section \ref{est}:}
\begin{equation}
\begin{aligned}
\grad{J^{\text{foc}}}{\phi_d}&=2\Im\left\{ \psi_0^*-\epsilon
  \psi_d\tf\left[\mathcal{M}\tf^{-1}\left(\psi_u \right)\right]^*\times
  \tf\left[\grad{J^{\text{foc}}}{h_c^{\text{foc}}}\left(\Psi_0-\epsilon\Psi_c\right)\right]\right\}\\
\grad{J^{\text{foc}}}{\phi_u}&=2\Im\left\{
  \psi_0^*\tf\left[\grad{J^{\text{foc}}}{h_c^{\text{foc}}}\left(\Psi_0-\epsilon\Psi_c\right)\right]
\right\} \\
&- \epsilon\psi_u^*\tf\left[\mathcal{M}^*\tf^{-1}\left(\Psi_d^*\tf\left\{\grad{J^{\text{foc}}}{h_c^{\text{foc}}}[\Psi_0-\epsilon\Psi_c]\right\}\right)\right]
\end{aligned}
\end{equation}

\review{with:}
\begin{equation}
\grad{J^{\text{foc}}}{h_c^{\text{foc}}}=
\frac{1}{{\sigma_n^{\text{foc}}}^2}[\alpha h_{\text{det}}(\alpha
h_{\text{det}}\star h_c^{\text{foc}}-i_c^\text{foc})]
\end{equation}
\review{and:}
\begin{equation}
\begin{aligned}
\psi_u&= P_ue^{j \phi_u}\\
\psi_d&= P_de^{j \phi_d}\qquad \Psi_d=\tf^{-1}(\psi_d)\\
\psi_0&= P_ue^{j(\phi_u+\phi_d)}\qquad \Psi_0=\tf^{-1}(\psi_0)\\
\Psi_c&= \tf^{-1}\{\psi_d\tf[\mathcal{M}\tf^{-1}(\psi_u)]\}\\
\end{aligned}
\end{equation}
\review{The regularization metric expression $\reg(\phi_k)$  ($k$ is for $u$ (upstream)
or $d$ (downstream )) is given by Eq. \eqref{eq_regf2}. Its gradient
$\grad{\reg}{\phi_k}$ can be written as:}
\begin{equation}
\grad{\reg}{\phi_k}=\mu_k\norm{\Delta\phi_k(r)}\text{.}
\end{equation}   
\review{where $\Delta$ represent the Laplacian operator.}

\end{document}